\documentclass[12pt]{article}

\usepackage{latexsym}

\begin{document}

\title{
Solar System Experiments and the Interpretation of Saa's Model of Gravity
with Propagating Torsion as a Theory with Variable Plank "Constant"}

\author{
      P. Fiziev \thanks{E-mail: fiziev@phys.uni-sofia.bg}\,
      \thanks{Permanent address: Department of Theoretical Physics,
                Faculty of Physics, Sofia University,
                5 James Bourchier Boulevard, Sofia~1164, Bulgaria}\\
{\footnotesize   Bogoliubov Laboratory of Theoretical Physics}\\
{\footnotesize   Joint Institute for Nuclear Research
  141980 Dubna, Moscow Region, Russia}\\
\\
      S. Yazadjiev \thanks{E-mail: yazad@phys.uni-sofia.bg}\\
{\footnotesize  Department of Theoretical Physics,
                Faculty of Physics, Sofia University,}\\
{\footnotesize  5 James Bourchier Boulevard, Sofia~1164, Bulgaria }\\
}

\maketitle

\begin{abstract}
It is shown that the recently proposed interpretation of the transposed
equi-affine theory of gravity as a theory with variable Plank "constant"
is inconsistent with basic solar system gravitational experiments.
\end{abstract}


\sloppy
\renewcommand{\baselinestretch}{1.3} %
\newcommand{\sla}[1]{{\hspace{1pt}/\!\!\!\hspace{-.5pt}#1\,\,\,}\!\!}
\newcommand{\db}{\,\,{\bar {}\!\!d}\!\,\hspace{0.5pt}}
\newcommand{\partb}{\,\,{\bar {}\!\!\!\partial}\!\,\hspace{0.5pt}}
\newcommand{\dsla}{\partb}
\newcommand{\eql}{e _{q \leftarrow x}}
\newcommand{\eqr}{e _{q \rightarrow x}}
\newcommand{\ite}{\int^{t}_{t_1}}
\newcommand{\itz}{\int^{t_2}_{t_1}}
\newcommand{\itd}{\int^{t_2}_{t}}
\newcommand{\lfrac}[2]{{#1}/{#2}}
\newcommand{\dV}{d^4V\!\!ol}
\newcommand{\ben}{\begin{eqnarray}}
\newcommand{\een}{\end{eqnarray}}
\newcommand{\la}{\label}

\bigskip

Recently a new model of gravity involving propagating torsion was proposed
by A. Saa \cite{Saa1}-\cite{Saa5}.
In this model a special type of Einstein-Cartan geometry is
considered in which the usual volume element $\sqrt{-g}\,d^4x$ is replaced
with new one:  $e^{-3\Theta}\sqrt{-g}\,d^4x$
-- covariantly constant with respect to
the transposed affine connection $\nabla^{T}$,
hence the name transposed-equi-affine theory of gravity \cite{F}.
As a result the torsion vector $S_\alpha = S_{\alpha\beta}{}^{\alpha}$
turns to be potential: $S_{\alpha}=
\partial_{\alpha}\Theta$, $\Theta$ being its scalar potential
\footnote{We use the Schouten's normalization conventions \cite{Schouten}
which differs from the original ones in \cite{Saa1}--\cite{Saa5}.}.

Because of the exponential factor $e^{-3\Theta}$ in the volume element
Saa's model has a very important feature: it leads to a consistent application
of the minimal coupling principle both in the action principle and in the
equations of motion for all matter fields.
These equations are of autoparallel type and
may be derived via the standard action principle for
{\em a nonstandard action integral}:
\ben
{\cal A}_{tot}= {\cal A}_G  +  {\cal A}_{M F} =
{\frac 1 c}\int \,{\cal L}_G \,e^{-3\Theta}\sqrt{-g}d^4x +
{\frac 1 c}\int \,{\cal L}_{M F} \,e^{-3\Theta}\sqrt{-g}d^4x,
\la{Atot}
\een
where ${\cal L}_G = - {\frac {c^2} {2\kappa}}R$ is the lagrangian of the
geometrical fields: the metric $g_{\alpha\beta}$,
and the torsion $S_{\alpha\beta}{}^{\gamma}$,
$R$ being the Cartan scalar curvature, $c$ being the speed of light,
and ${\cal L}_{M F}$ is the usual lagrangian of the corresponding
matter fields: scalar fields $\phi(x)$, spinor fields $\psi(x)$,
electromagnetic fields $A_\alpha(x)$,
Yang-Mills fields ${\bf A}_\alpha(x)$, e.t.c.

But this property not held for the equations of motion of classical particles
and fluids which turn to be of geodesic type \cite{F}. Most probably this
inconsistency leads to the negative result obtained in \cite{FY}:
Saa's model is inconsistent with the basic solar system
gravitational experiments.

Then having in mind to preserve the good features of Saa's model
and in the same time somehow to avoid this problem
we are forced to try some further modifications of the model.
The simplest one is to use the Saa's modification of the volume element
{\em only in the action integrals} like (\ref{Atot})
and the usual volume element in all other physical,
or geometrical formulae \cite{F}.
This leads to the action for a classical spinles particle in a form:
\ben
{\cal A}_m = - mc\int e^{-3\Theta}ds
\la{ACPart}
\een
where $m$ is the rest mass of the particle and
$ds=\sqrt{g_{\alpha\beta} dx^\alpha dx^\beta}$ is the usual
four-dimensional interval.
The corresponding action integral for spinless fluid (See for details \cite{F})
is:
\ben
{\cal A}_\mu ={\frac 1 c}\int \,{\cal L}_\mu\,e^{-3\Theta}\sqrt{-g}d^4x
=-{\frac 1 c}\int \,(\mu c^2+\mu \Pi)\,e^{-3\Theta}\sqrt{-g}d^4x,
\la{ACFlu}
\een
$\mu(x)$ being fluid's density,
$\Pi$ being the elastic potential energy of the fluid.

This situation calls for a new curious interpretation
of the torsion potential $\Theta$ as a quantity which describes
the space-time variations of the Plank "constant'' according to the law
\ben
\hbar (x) = \hbar_\infty  e^{3\Theta(x)},
\la{Plank}
\een
$\hbar_\infty$ being the Plank constant in vacuum far from matter.

Indeed, according to the first principles
we actually need lagrangians and action integrals to write down
the quantum transition amplitude in a form of Feynman path integral on the
histories of all fields and particles.
In the variant of the theory under consideration it has the form:
\ben
\int{\cal D}\left(
\vbox to 12pt{}g_{\alpha\beta}(x), S_{\alpha\beta}{}^{\gamma}(x),
\phi(x),\psi(x), A_\alpha(x), {\bf A}_\alpha(x),
...;x(t),...\right) \nonumber \\
\exp\left( {\frac 1 {\hbar_\infty}}\left(\int d^4x e^{-3\Theta(x)}
(L_G + L_{M F}) - mc\int e^{-3\Theta}ds \right) \right).
\la{QA}
\een

Now it is obvious that the very Plank constant $\hbar$ may be
included in the factor $e^{3\Theta(x)}$, but more important is the
observation that we must do this, because the presence of this {\em uniform}
factor in the formula (\ref{QA}) means that we actually introduce a local
Plank "constant'' at each point of the space-time.
Indeed, if the geometric field $\Theta(x)$ changes slowly in a cosmic scales,
then in the framework of the small domain of the laboratory we will see
an effective "constant":
$\hbar (x)\approx\hbar_\infty e^{3\Theta(x_{laboratory})}= const=\hbar$.

In presence of  spinless matter only an Einstein-Cartan geometry with
semi-symmetric torsion tensor
$S_{\alpha\beta}{}^{\gamma}=S_{[\alpha}\delta_{\beta]}^{\gamma}$
appears and the following equations for geometrical fields are obtained
\ben
G_{\mu\nu} +\nabla_\mu \nabla_\nu \Theta - g_{\mu\nu}\Box\Theta =
{\frac \kappa {c^2}}
\left((\varepsilon + p)u_{\mu}u_{\nu} - pg_{\mu\nu} \right)\nonumber\\
\nabla_\sigma S^\sigma= \Box\Theta= -{\frac {2\kappa} {c^2}}
\left(\varepsilon + 3p\right)
\la{SYS}
\een
where  $G_{\mu\nu}$ is Einstein tensor with respect to the affine connection
$\nabla_{\mu}$, $\kappa$ being the Einstein gravitational constant,
$\varepsilon$, $p$ and $u_{\mu}$ are the energy density, pressure
and four velocity of the relativistic perfect fluid \cite{F}.
Using the standard  variational principle for the action (\ref{ACFlu})
one can obtain the equations of motion for the perfect fluid:
\ben
(\varepsilon + p)u^{\beta}\nabla_{\beta}u_{\alpha}=
\left(\delta_{\alpha}^{\beta} - u_{\alpha}u^{\beta}\right)\nabla_{\beta}p
+ \cal F_{\alpha}
\la{FEM}
\een
where $${\cal F_{\alpha}}= -2(\varepsilon + p)
\left(\delta_{\alpha}^{\beta} - u_{\alpha}u^{\beta}\right)\nabla_{\beta}
\Theta$$ is the torsion force, as defined in \cite{F}.

This nonzero value of the torsion force shows that in the present model
with variable Plank "constant" (VPC model) the matter equations
of motion are not of autoparallel, nor of geodesic type in contrast to
all equations for matter fields which are of autoparallel type.
This inconsistency of the model is not enough to reject it immediately
as far as the very requirement for all dynamical equations in theory to be
of the same type is not founded on a well established principle,
nevertheless it seems to be necessary for validity of the corresponding
generalization of the equivalence principle in spaces with torsion \cite{F2}.

The main purpose of this letter is to investigate the consistency of
the VPC model with basic solar-system experimental facts.
To do this we have to consider the motion of a test particle
in presence of a metric and torsion fields.
The standard variation of the action (\ref{ACPart}) yields the equations
of motion we need, but it's more convenient to investigate directly
the corresponding Hamilton-Jacoby equation:
\ben
g^{\mu\nu}\partial_{\mu}S\partial_{\nu}S =\left( mce^{-3\Theta}\right)^2.
\la{HJ}
\een
The conform transformation
$g_{\mu\nu} \rightarrow \,\,\stackrel{*}{g}_{\mu\nu}= e^{-6\Theta}g_{\mu\nu}$
yields the effective metric $\stackrel{*}{g}_{\mu\nu}$ and
the following form of the equation (\ref{HJ})
\ben
\stackrel{*}{g}^{\mu\nu}\partial_{\mu}S\partial_{\nu}S =m^2c^2
\een
which is well known from general relativity.
Thus we may consider the motion of a test particles precisely as in
general relativity working with the metric  $\stackrel{*}{g}_{\mu\nu}$.
Therefore the simplest way to compare the predictions of the VPC model with
the experimental facts is to consider post-Newtonian expansion of the metric
$\stackrel{*}{g}_{\mu\nu}$ in vacuum in vicinity of a star like the Sun.

The asymptotically flat, static and spherically symmetric general solution
of the equations (\ref{SYS}) for geometric fields in vacuum is known
\cite{Brans}, \cite{ZX}. In isotropic coordinates it's given by a
two-parameter -- $(r_{0},k)$  family of solutions
\ben
ds^2 =
\left(1-{r_{0}\over r} \over 1 +{r_{0}\over r}\right)^{2\over \rho(k)}(c\,dt)^2
- \left(1 - {r_{0}^2 \over r^2}\right)^2
\left(1-{r_{0}\over r} \over 1+{r_{0}\over r}\right)^{{2\over \rho(k)}(3k - 1)}
\left(dr^2  + r^2d\Omega^2\right) ,\\
\Theta={k\over 2}\nu
\la{metric}
\een
where $\rho(k)=\sqrt{3\left(k - {1\over 2}\right)^2 + {1\over 4}}$.
In the VPC model under consideration the whole geometry (metric and torsion)
causes a gravitational force (of pure geometrical nature).
The parameter $k$ presents the ratio of the torsion part of this force
and its metric part.
In the case when $k = 0$ we have the usual torsionless Schwarzshild's
solution and $r_g \equiv 4r_0$ is the standard gravitational radius.

From  equations (\ref{metric}) we obtain the effective metric
$\stackrel{*} g_{\mu\nu}$ and the effective four-interval
\ben
d\stackrel{*} s^2 =
\left(1 - {r_{0}\over r} \over 1 +{r_{0}\over r}\right)^
{{2\over \rho(k)}(1 - 3k)}(c\,dt)^2
- \left(1 - {r_{0}^2 \over r^2}\right)^2
\left(1-{r_{0}\over r} \over 1+{r_{0}\over r}\right)^
{{-2\over \rho(k)}}
\left(dr^2  + r^2d\Omega^2\right).
\la{efmet}
\een

The asymptotic expansion of the metric in (\ref{efmet}) at
$r\rightarrow \infty$ gives
\ben
d\stackrel{*}s^2 \approx \left(1 - {4r_{0}(1 - 3k)\over \rho(k)r} +
{8r_{0}^2(1 - 3k)^2\over \rho(k)^2 r^2}\right)(c\,dt)^2
- \left(1 + {4r_{0}\over \rho(k)r}\right)
\left(dr^2  + r^2d\Omega^2\right) .
\la{As1}
\een
In the asymptotic region $r\rightarrow \infty$ we must have Newtonian gravity.
Consequently the mass "seen" by the test particles is
\ben
M = {2r_{0}(1 - 3k)\over \rho(k)}
\la{Mass}.
\een
Therefore we may represent the effective four-interval in the asymptotic form
\ben
d\stackrel{*} s^2 \approx
\left(1 - {2M \over r} + {2M^2\over r}\right)(c\,dt)^2 -
\left(1 + {1\over 1- 3k} {2M\over r}\right)
\left(dr^2 + r^2d\Omega^2 \right).
\een
From the above expression it immediately follows that two of post-Newtonian
parameters corresponding to the effective metric  $\stackrel{*} g_{\mu\nu}$ are
\ben
\stackrel{*}  \beta = 1 , \,\,\,\,\,\,\, \stackrel{*}  \gamma = {1\over 1-3k}.
\een
As it's well known, solar system gravitational  experiments set tight
constrains on post-Newtonian parameters \cite{Will}:
\ben
\mid  \stackrel{*}  \beta  - 1 \mid  <1*10^{-3},\,\,\,\,\,\,
\mid  \stackrel{*}  \gamma - 1 \mid  <2*10^{-3}.
\een
Therefore, to avoid contradictions with the basic experimental facts we
must have
\ben
\left| {3k \over 1 - 3k} \right| < 2*10^{-3}.
\la{CondK}
\een
In order to specify the theoretically possible values of $k$
we must investigate a model of a star. As a simplest basic model we may
consider a static spherically symmetric star.
Putting the metric in the standard form
$$ds^2 = e^{\nu}(c\,dt)^2 - e^{\lambda}dr^2 - r^2(d\theta^2 +
\sin^2(\theta)d\varphi^2)$$
we obtain from the general field equations (\ref{SYS}), (\ref{FEM})
the following complete system of ordinary differential equations for
the star's fluid equilibrium
\ben
\xi^{\prime} + {2\over r}\xi + \left(2\xi - \lambda^{\prime}\over 2\right)
\xi - 3S_{r}\xi = -(\varepsilon + 3p)e^{\lambda} \nonumber  \\
S_{r}^\prime + {2\over r}S_{r} + \left(2\xi - \lambda^{\prime}\over 2\right)
S_{r} - 3S_{r}^2 = -(\varepsilon + 3p)e^{\lambda}  \nonumber \\
e^{\lambda} - \left(1 + {r\over 2}(2\xi - \lambda^\prime)\right) =
-3rS_{r} + 2(\varepsilon + 2p)r^2 e^{\lambda} \nonumber \\
\xi^\prime - {\lambda^\prime\over 2}\xi + {\xi}^2 - {\lambda^\prime\over r}=
3S_{r}^\prime - {3\over 2}\lambda^\prime S_{r} + 3S_{r}^2
-2(\varepsilon + 2p)e^{\lambda} \nonumber \\
p^\prime = - (\varepsilon + p)(\xi - 3S_{r}) \nonumber \\
p = p(\varepsilon)
\la{stareq}
\een
where $\xi ={1\over 2} \nu^\prime$ , $S_{r}= \Theta^\prime$,
$p = p(\varepsilon)$ is the matter state equation, $\varepsilon$
and $p$ are the energy density and the pressure.
The prime denotes differentiation with respect to $r$.

The regular at the center of the star ($r=0$) solution corresponds to
the initial conditions \cite{BFY}:
$$ \xi(0)=0, \,\,\,S_{r}(0)=0. $$
As we see the first two equations of the system (\ref{stareq}) coincide.
Then by virtue of the same initial conditions for $\xi$ and $S_{r}$
we obtain equal solutions $\xi = S_{r}$. Hence, in VPC model the only possible
value of the parameter $k$ is $k = 1$. This means that in this model the
torsion part of gravitational force equals to the metric one in magnitude.
As a consequence it is impossible to fulfill the condition (\ref{CondK}).
Moreover, if $r_0 > 0$ the value $k=1$ leads to a negative mass of the star
(See equation (\ref{Mass})).

This result shows that the interpretation of the Saa's model
as a theory with variable Plank's constant is inconsistent with
the well known solar system gravitational experiments.

\newpage

\noindent{\Large\bf Acknowledgments}
\bigskip

This work has been partially supported by
the Sofia University Foundation for Scientific Researches, Contract~No.~245/98,
and by
the Bulgarian National Foundation for Scientific Researches, Contract~F610/98.

One of us (PF) is grateful to the leadership of the
Bogoliubov Laboratory of Theoretical Physics, JINR, Dubna, Russia
for hospitality and working conditions during his stay there in
the summer of 1998.

\end{document}